\documentclass[preprintnumbers, pra, showpacs, floatfix,twocolumn,
preprintnumbers, letterpaper, superscriptaddress]{revtex4}
\usepackage{amsfonts}
\usepackage{amsmath}
\usepackage{amssymb,epsf}
\usepackage{latexsym}
\usepackage{graphicx,epsfig}
\usepackage{amssymb}
\usepackage{subfigure}
\usepackage[colorlinks=true,citecolor=blue,linkcolor=blue,urlcolor=black]{hyperref}
\usepackage{epstopdf}
\usepackage{color}

\def\be{\begin{equation}}
\def\ee{\end{equation}}
\def\ba{\begin{eqnarray}}
\def\ea{\end{eqnarray}}
\def\la{\langle}
\def\ra{\rangle}

\begin{document}
\title{ Ising order parameter and topological phase transitions:\\ Toric code in a uniform magnetic field}
\author{Mohammad Hossein Zarei}
\email{mzarei92@shirazu.ac.ir}
\affiliation{Physics Department, College of Sciences, Shiraz University, Shiraz 71454, Iran}
\begin{abstract}
Quantum Ising model in a transverse field is of the simplest quantum many-body systems used for studying universal properties of quantum phase transitions. Interestingly, it is well-known that such phase transitions can be mapped to topological phase transitions in Toric code models. Therefore, one can expect that well-known properties of the transverse Ising model are used for characterizing topological phase transition in Toric code model. In this paper, we consider the magnetization of Ising model and show while it is a local order parameter, it is mapped to a non-local order parameter in the topological side. Consequently, we introduce a string order parameter for topological phase transitions in Toric code models defined on different lattices with different dimensions in presence of a uniform magnetic field. Since such string order parameter is dual of the Ising order parameter with well-known properties, it leads to a topological (non-local) characterization of phase transition in the Toric code model in uniform field. Our results show that although topological phase transitions do not follow a symmetry-breaking mechanism, it might be still possible to use concepts of symmetry-breaking phase transitions for topological ones by finding suitable mappings.
\end{abstract}
\maketitle
\section{Introduction}
One of the most important problems in condensed matter physics is to characterize different phases of matter where concept of the phase transition has emerged \cite{golden}. Specifically, symmetry-breaking theory of Landau has been known as the most well-established paradigm used to classify different phases based on different symmetries\cite{landa}. However, it has been shown there is still a big zoo of phases which do not follow the Landau paradigm where there are different phases with the same symmetries \cite{3}. These new phases have topological order with a non-local nature where they can not be characterized by any local order parameter\cite{4,5,6,7,8,9,10,11}. Therefore, introducing a comprehensive mathematical frame as well as a physical mechanism for characterizing and classifying topological phases has remained a challenging and open problem\cite{stringnet,wwen,wen3}.

Topological order has also been attracted much attention in quantum information science where quantum systems with topological order have been one of the most important candidates for quantum storage and quantum computation \cite{del,18,19,20,21,22,23,24}. Of the most interesting topological systems are topological quantum codes such as Toric code (TC) which has been introduced in context of quantum error correction \cite{26}. It has been shown that non-local nature of such quantum codes leads to a natural robustness against local perturbations \cite{27,28,schmit}. Beside the above practical applications, topological quantum codes can also be used as toy models for studying topological order because of their simplicity where topological phase transition out of a topological code state has attracted much attention\cite{jahromi2013, kargar2013, karimipour, zarei2016,29,zareimon}.

On the other hand, an important step toward characterizing topological phase transitions is to introduce measures which can capture non-local nature of the transition. Such measures are usually some topological quantum numbers such as topological degeneracy and topological entanglement entropy \cite{15,16,13} which do not behave like a traditional order parameter. However, in comparison with ordinary quantum phase transitions one can expect there is also an order parameter with a non-local nature for a topological phase transition \cite{fen,17,32,34,38}. Interestingly, there are mappings between topological phase transitions and ordinary quantum phase transitions which have been introduced for lattice gauge theories in \cite{l1,l2} and also for Calderbank-Stean-Shor (CSS) code models in \cite{29}, see also \cite{zareimond} for a mapping to classical phase transitions.  By such mappings, one can expect that the well-established knowledge of quantum phase transitions are used for understanding their corresponding topological phase transitions.

One of the most well-known quantum phase transitions are transition in transverse Ising models \cite{tising,trans}. They are in fact the best models for introducing symmetry-breaking mechanism. A symmetry-breaking phase transition in the transverse Ising model is characterized by a simple local order parameter namely magnetization and different universality classes can be found. In this regard and since TC model in a uniform magnetic field is mapped to the transverse Ising model, One will be able to study topological phase transition in TC model by using the well-established knowledge of the transverse Ising model \cite{Sm1,Sm2,Sm3,Sm4}. In this paper, we take an important step in this direction by considering Ising order parameter in the transverse Ising model. We show that this quantity is mapped to a non-local order parameter for TC in a uniform field. Therefore, we introduce a string order parameter for topological phase transition from TC state to a magnetized state. Our result is held for TC's defined on any graphs in any dimension.

In Sec.(\ref{sec1}), we define TC model in a uniform magnetic field where we give a brief review on TC and its topological order. In Sec.(\ref{sec2}), we show how TC in an unifrom field is mapped to a transverse Ising model by a change of basis from a computational basis to a non-local basis. Finally, in Sec.(\ref{sec3}), we introduce string order parameter which is dual of Ising order parameter. Therefore, we characterize topological phase transition in TC in a uniform field by a non-local order parameter.

\section{Toric code models in presence of a uniform magnetic field}\label{sec1}
In this section, we give an introduction to TC models in presence of a uniform magnetic field. To this end, let us start with the original version of TC which is defined on a two-dimensional (2D) square lattice and then we will explain how it can be extended to other lattices in different dimensions.

\begin{figure}[t]
\centering
\includegraphics[width=9cm,height=6cm,angle=0]{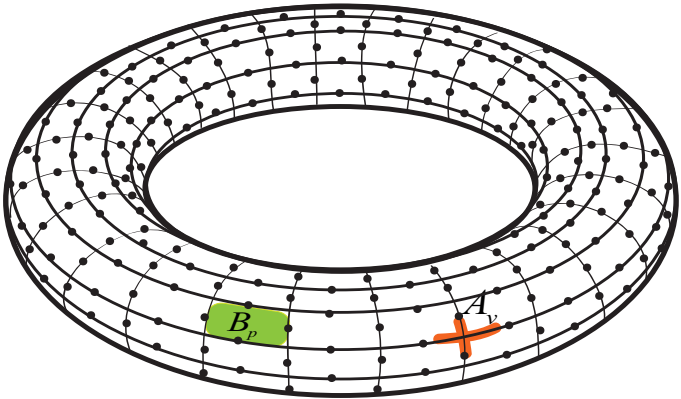}
\caption{(Color online) A two dimensional square lattice attached on a torus with qubits living in edges. Plaquette and vertex operators i. e. $B_p$, $A_v$ are defined corresponding to plaquettes and vertices, respectively. } \label{k1}
\end{figure}

As it has been shown in Fig.(\ref{k1}), consider an $L\times L$ 2D square lattice attached to a torus with qubits living in edges where we call them edge qubits and denote them by $e$. Corresponding to each plaquette and vertex of the lattice, two kinds of operators are defined in the following form:
\begin{equation}\label{sta}
B_p =\prod_{e\in \partial p} Z_e ~~,~~A_v =\prod_{e\in v}X_e
\end{equation}
where $p$ refers to a plaquette of the graph and $e\in \partial p$ refers to all edge qubits around a plaquette $p$. $e\in v$ refers to all edge qubits incoming to vertex $v$ and $Z$ and $X$ are Pauli operators. These operators belong to the Pauli group $P_n$ and since they commute with each other, they technically are called stabilizer operators which generate a stabilizer group \cite{gottesman}. Then, the TC in a uniform magnetic field is defined by the following Hamiltonian:
\begin{equation}\label{TM}
H=-J\sum_v A_v -J\sum_p B_p -h\sum_e Z_e
\end{equation}
where $J$ refers to coupling of TC and $h$ refers to a single unifrom magnetic field in $z$ direction. Here, we have considered the above simple form because, as we will show in the next section, there is a simple dual mapping for this specific case while there is not such duality for a general uniform magnetic field.

Turning to Eq.(\ref{TM}), it is clear that in limit of $h/J \rightarrow \infty$, the ground state of the above Hamiltonian will be a magnetized state of $|00...0\ra$ where $|0\ra$ refers to positive eigenstate of the $Z$ operator. Interesting properties of the above Hamiltonian reveals in limit of $h/J \rightarrow 0$ where the Hamiltonian will describe the TC with a topological degeneracy that we explain as follows. One of the ground states can be easily found in the following unnormalized form:
\begin{equation}\label{kitaevstate}
|\psi_k \ra =\prod_{v}(1+A_v)|0\ra ^{\otimes n}
\end{equation}

\begin{figure}[t]
\centering
\includegraphics[width=7cm,height=7cm,angle=0]{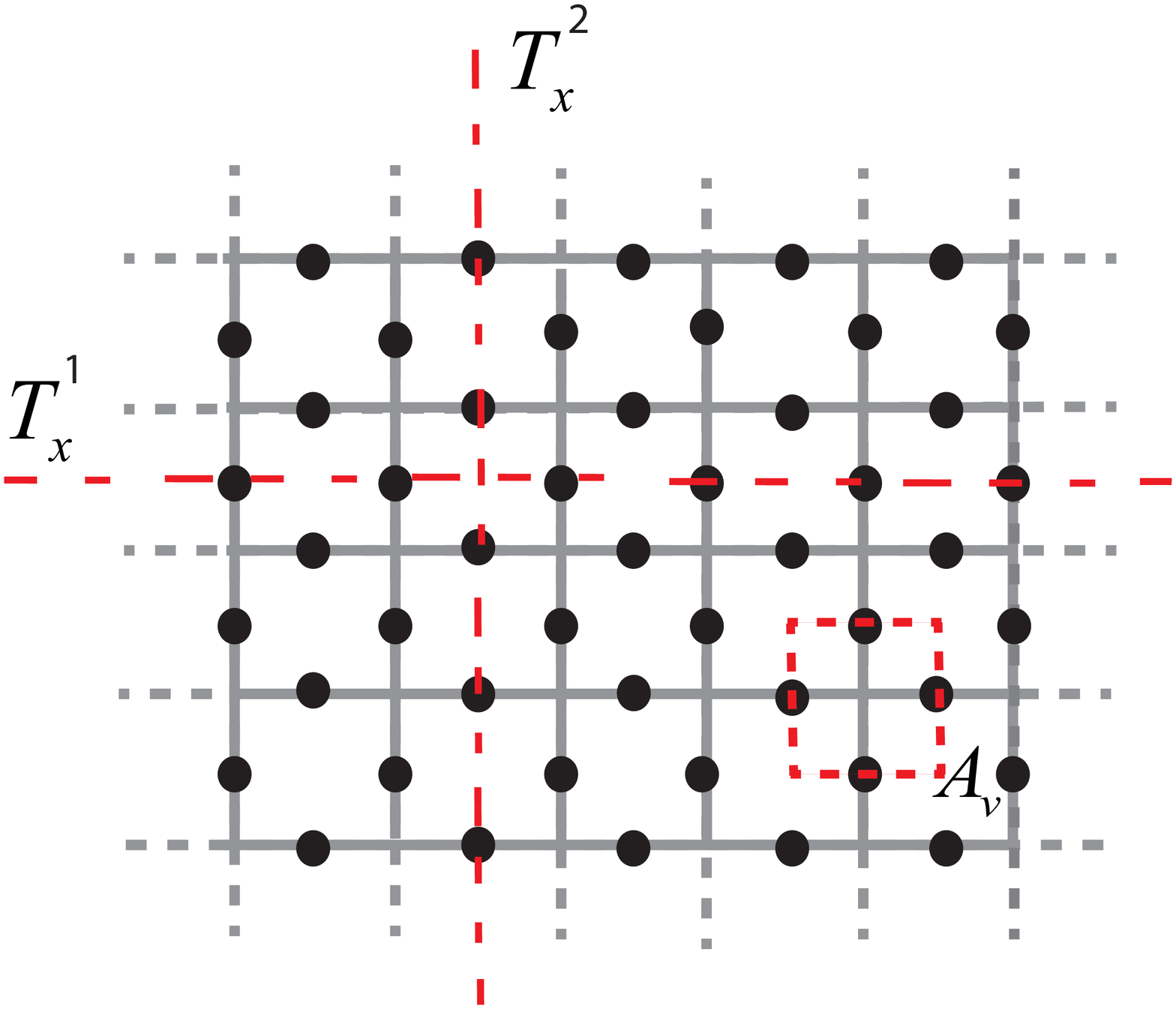}
\caption{(Color online) A vertex operator $A_v$ is represented by a loop in dual lattice. There are two kinds of non-trivial loop operators i. e. $T^1 _x$, $T^2 _x$ turning around the torus in two different directions on dual lattice. Non-trivial loops are not boundary of a surface on the torus and therefore, non-trivial loop operators can not be constructed by a product of $A_v$'s.} \label{kitaev}
\end{figure}

where $n=2L^2$ is the number of qubits. Since operators $A_v$ can be represented by loops in the dual lattice, see Fig.(\ref{kitaev}), the above state can be interpreted by a superposition of loops $1$ in a sea of $0$ which is called loop-condensation. Furthermore, because of topology of torus, there are also two different loop operators which can not produced by a product of $A_v$'s. As it is shown in Fig(\ref{kitaev}), we denote them by $T^1 _x$ and $T^2 _x$ and other ground states of the TC can be generated by them in the following form:
\begin{equation}\label{deg}
|\psi_{\mu , \nu}\ra=(T^1 _x)^\mu(T^2 _x)^\nu |\psi_k \ra
\end{equation}
\begin{figure}[t]
\centering
\includegraphics[width=3cm,height=3cm,angle=0]{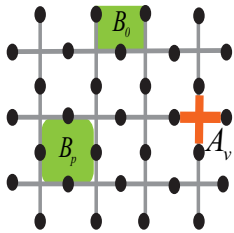}
\caption{(Color online) TC can be defined on a 2D lattice with open boundary condition. Plaquette operators corresponding to the boundary of the lattice must be three-local.} \label{k2}
\end{figure}
where $\mu , \nu =\{0,1\}$ and therefore there is a four-fold degenerate ground subspace for the TC. An important point here is that such a degeneracy does not depend on symmetries of the Hamiltonian but it is a result of the periodic boundary condition. For example, one can define the TC on a 2D square lattice with open boundary conditions. As it is shown in Fig.(\ref{k2}), vertex and plaquette operators are defined as before but plaquette operators in boundary are three-local. such a model does not have any degeneracy and the unique ground state is in the form of Eq.(\ref{kitaevstate}). Dependence of degeneracy on the boundary condition is an important property of topological orders. Furthermore, topological nature of degenerate ground space can be revealed by non-local order hidden in four degenerate ground states of the TC.
\begin{figure}[t]
\centering
\includegraphics[width=7cm,height=7cm,angle=0]{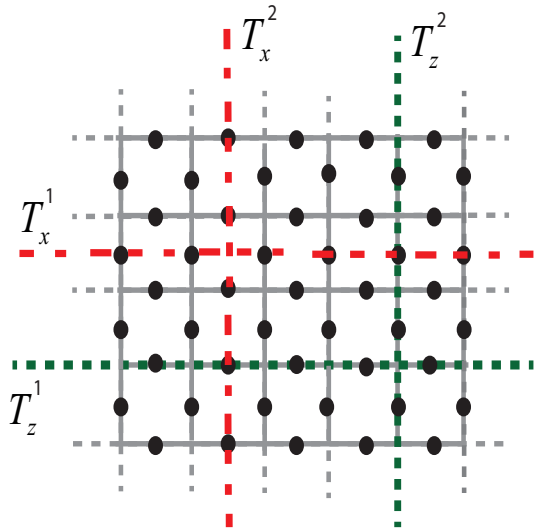}
\caption{(Color online) Corresponding to non-trivial loops on the lattice, there are two kinds of $Z$-type loop operators i. e. $T^1 _z$, $T^2 _z$ which cross non-trivial loops corresponding to operators of $T^1 _x$ and $T^2 _x$ for one time. } \label{kitaev2}
\end{figure}
In other words, there is no local order parameter which can distinguish these states from each other. As it is shown in Fig.(\ref{kitaev2}), there are two non-local parameters which are constructed by a product of $Z$ operators around two direction of the torus. We call them $T^1 _z$ and $T^2 _z$ and it is simple to check that they have the following anti-commutation relations with operators $T^1 _x$ and $T^2 _x$:
\begin{equation}
\{T^1 _x , T^2 _z\}=0~~,~~\{T^2 _x , T^1 _z\}=0
\end{equation}
Therefore, it is clear that the expectation values of $T^1 _z$ and $T^2 _z$ for four states of (\ref{deg}) will be different and these two operators can distinguish them from each other.

\begin{figure}[t]
\centering
\includegraphics[width=5cm,height=10cm,angle=0]{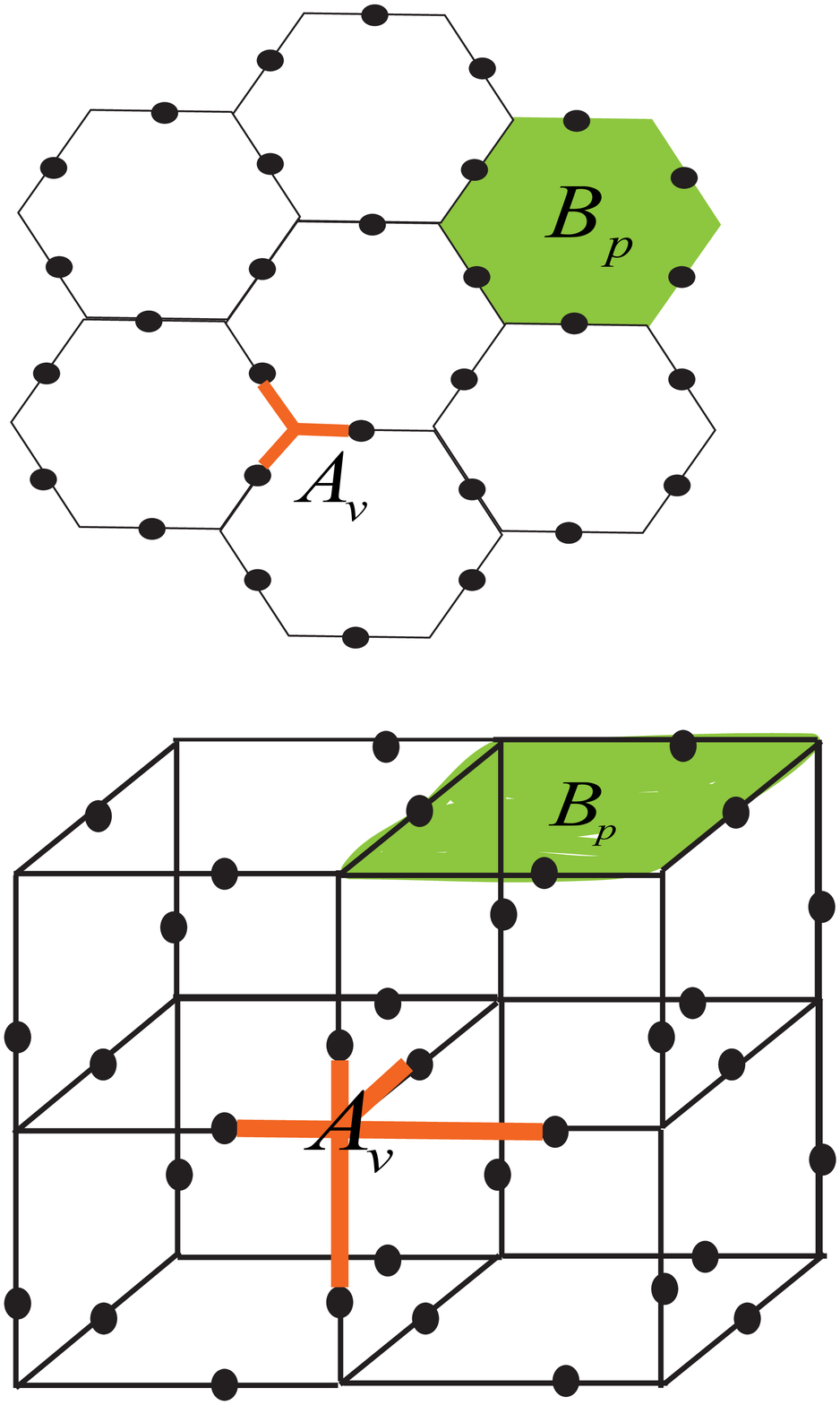}
\caption{(Color online) TC can be defined on other lattices. Plaquette and vertex operators are defined in a similar way.} \label{3d}
\end{figure}

In this way, the Hamiltonian (\ref{TM}) involves two completely different phases in two limits of magnetic field where the ground state is a topological state with a four fold degeneracy at $h/J \rightarrow 0$ and is a trivial state at $h/J \rightarrow \infty$. Therefore, there will be a topological phase transition in a critical value of $(h/J)^*$ what we will consider in the next section. Before that, we would like here to emphasize that TC can be also defined on other lattices in different dimensions. For example, in Fig.(\ref{3d}) we have shown a two dimensional hexagonal lattice and a three dimensional cubic lattice. Vertex and plaquette operators are defined in the same way and the ground state will have the same form of Eq.(\ref{kitaevstate}) but degeneraacy depends on the boundary condition. Another important point is that the above topological phase transition happens for all boundary conditions and the topological nature of the transition does not depend on the degeneracy of the ground state. Therefore, we can use a suitable boundary condition so that our calculations can be easily done.
\section{Mapping to quantum Ising model}\label{sec2}
\begin{figure}[t]
\centering
\includegraphics[width=5cm,height=5cm,angle=0]{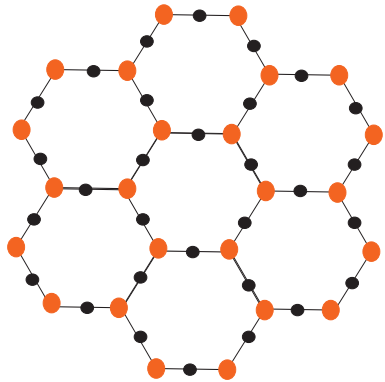}
\caption{(Color online) Initial Hamiltonian is written in the computational basis corresponding to edge qubits. The dual basis can be represented by qubits living in the vertices denoted by red (light) circles. } \label{verq}
\end{figure}

In this section, we want to consider topological phase transition from TC state to a magnetized state. Instead of finding an exact form for the ground state of the Hamiltonian of (\ref{TM}), here we perform a change of basis to map the above Hamiltonian to a quantum Ising model in a transverse field. Our mapping is independent of the graph in which the TC has been defined. In other words, we show that the TC in a uniform magnetic field defined in an arbitrary graph $G$, is mapped to a transverse Ising model defined on the same graph $G$. \\
Consider an arbitrary graph $G=(V,E)$ where $V$ refers to set of vertices and $E$ refers to set of edges. Qubits live in edges which called edge qubits, vertex and plaquette operators are defined corresponding to each vertex and plaquette of the graph in the same form of (\ref{sta}). Next, in order to map the Hamiltonian (\ref{TM}) to the transverse Ising model, we introduce a new basis, which is called dual basis, for rewriting the Hamiltonian (\ref{TM}). We insert new qubits in each vertex of the graph called vertex qubits, see Fig.(\ref{verq}), and we define the following basis corresponding to the vertex qubits:
\begin{equation}\label{bas}
|r_1 , r_2 , ... , r_m \rangle =\prod_{i}(1+(-1)^{r_i} A_i)|0\ra ^{\otimes n}
\end{equation}
where we have used numbers of $1,2,..m$ for denoting vertices where $m$ is the total number of vertices and $|r_i\ra$ refers to the state of a vertex qubit inserting in vertex of $i$. We have also ignored normalization factor in the above basis. Furthermore, it is clear that the above basis in not a complete basis. In fact, we can also write other bases generated by $B_p$ operators in the form of $ \prod_{p}(1+(-1)^{r'_p} B_p)|+\ra ^{\otimes n}$ where $|+\ra$ refers to the positive eigenstate of the pauli operator $X$. However since operators $Z_e$ commute with $B_p$, the effect of magnetic term in this basis is trivial. Therefore, we only rewrite the Hamiltonian (\ref{TM}) in the dual basis of (\ref{bas}). Next, we should find the form of different terms in the Hamiltonian (\ref{TM}) when they are applied in the basis of (\ref{bas}).

Let us start with the operator of $A_i$. Since $A_i ^2 =1$, it is concluded that $ A_i(1+(-1)^{r_i} A_i)=(-1)^{r_i}(1+(-1)^{r_i} A_i)$. Therefore, the effect of $A_i$ in the dual basis is in the form of $A_i |r_1 , r_2 , ... , r_m \rangle =(-1)^{r_i}|r_1 , r_2 , ... , r_m \rangle$. It means that the operator $A_i$ in the dual basis is the same as the Pauli operator $\sigma_z$ and we denote it by the logical form of $\bar{Z}_i$.

Next, we consider the effect of magnetic term in the dual basis where the operator $Z_e$ should be applied. As it is shown in Fig.(\ref{isi}-a),  consider an edge qubit $e$ in the graph belonging to two vertices of the graph denoted by $A_1$ and $A_2$. It is clear that the corresponding $Z_e$ operator commutes with all $A_i$'s instead of two vertex operators $A_1$ and $A_2$ corresponding to two vertices in two endpoints of the $e$. Consequently, when we apply the operator $Z_e$ to the dual basis, we will have $
Z_e \prod_{i}(1+(-1)^{r_i} A_i)|0\ra ^{\otimes n}=(1+(-1)^{r_1 +1} A_1)(1+(-1)^{r_2 +1} A_2)\prod_{i\neq \{1,2\}}(1+(-1)^{r_i} A_i)Z_e |0\ra ^{\otimes n}$.
On the other hand, since $Z_e |0\ra =|0\ra$  we will have:
\begin{figure}[t]
\centering
\includegraphics[width=6cm,height=3cm,angle=0]{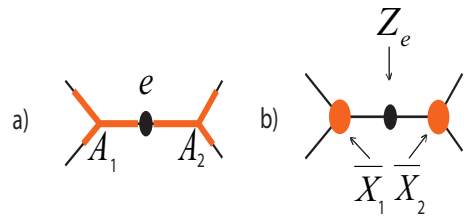}
\caption{(Color online) a) An edge qubit of $e$ belongs to two vertices of $1$ and $2$. Therefore, a $Z_e$ operator does not commute with vertex operators $A_1$ and $A_2$. b) Effect of a $Z_e$ operator on a edge qubit in the computational basis is equal to effect of an $\bar{X}_1 \bar{X}_2$ operator on vertex qubits in the dual basis.}\label{isi}
\end{figure}
\begin{equation}
Z_e |r_1 , r_2 ,..., r_m \ra =|r_1 +1 ,r_2 +1 ,..., r_m\ra
\end{equation}
It means that the operator $Z_e$ behaves like the flip operator on vertex qubits $1, 2$ and therefore it is the same as operator of $\bar{X}_1 \bar{X}_2$ in the dual basis, see Fig.(\ref{isi}-b). Since this argument is held for all $Z_e$ operators in the Hamiltonian (\ref{TM}), it is concluded that the magnetic term in the original Hamiltonian is mapped to Ising interaction between neighbor vertex qubits.

In this way, the Hamiltonian (\ref{TM}) in the dual basis will have the following dual form which is a quantum Ising model in a transverse field on the graph $G$:
\begin{equation}\label{dual}
\bar{H}=-J\sum_i \bar{Z}_i -h\sum_{\la i,j\ra}\bar{X}_i \bar{X}_j -C
\end{equation}
where the constant of $C$ refers to trivial effect of operators of $B_p$ in the dual basis in a sense that the above dual mapping only holds for the subspace where all $B_p$ operators are in the ground state and the anyon statistics plays no role. Interestingly, the above dual Hamiltonian is a quantum Ising model in a transverse field which has been defined in the same original graph $G$ with qubits living in vertices. Furthermore, coupling constant and magnetic field in the TC model have been converted to magnetic field and coupling constant in the Ising model, respectively. Therefore, it is a kind of the strong-weak coupling duality \cite{29}. We emphasize again that the above mapping was independent of the kind of graph and therefore, a TC model in uniform field on any arbitrary graph with qubits living in edges is mapped to a transverse Ising model on the same graph with qubits living in the vertices. However, there is a point related to the boundary condition where it might lead to some different terms in the dual Hamiltonian corresponding to the boundary condition of the initial Hamiltonian. We will explicitly explain this point for a specific boundary condition in the next section.

Finally note that while there is a topological phase transition in the TC model in uniform field, the transverse Ising model shows a symmetry-breaking phase transition. It is important because the former have a non-local nature while the latter is characterized by a local order parameter. The existence of such a mapping is in fact a consequence of non-local nature of the dual basis. In other words, since we have rewritten the initial Hamiltonian in a non-local (topological) basis, topological properties of the model are encoded in the dual basis and the final Hamiltonian has found a local property. In spite of such a deep transformation, the final Hamiltonian have the same energy spectrum with the initial Hamiltonian, because our transformation is an unitary one. Specifically, the quantum phase transition point is the same for both the above models. However, one can ask if there is another point beyond the phase transition point. In particular interesting question is that if such a mapping can help to characterize the topological phase transition in TC by using symmetry-breaking mechanism in Ising model. In the next section, we take a step in this direction, where we start with Ising order parameter in Ising model and find a string order parameter for TC where it characterizes the topological phase transition.

\section{Ising order parameter and a string order parameter for TC}\label{sec3}
Quantum phase transition in the transverse Ising model is of the most well-known problems in quantum many-body systems. Specifically, such a transition can be characterized by a local order parameter which is in fact the magnetization and is equal to expectation value of Pauli operator $X$ on each qubits $i.~e.~\la X_i \ra$. It is well-known that this quantity in terms of ratio of magnetic field and Ising coupling shows a continuous (second-order) phase transition in a critical value which can be found by numerical calculations for different lattices \cite{trans}.
\begin{figure}[t]
\centering
\includegraphics[width=5cm,height=5cm,angle=0]{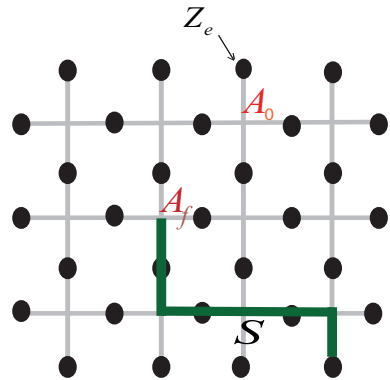}
\caption{(Color online) For TC with an open boundary, A $Z_e$ operator in the boundary does not commute with the vertex operator $A_0$. Therefore, it is equal to an $\bar{X}_0$ in the dual basis. A $Z$-type string operator corresponding to a green (dark) string of $S$  with endpoints living in the boundary and  in vertex of $f$ does not commute with a vertex operator of $A_f$. Therefore, it is equal to an $\bar{X}_f$ operator in the dual basis.} \label{string}
\end{figure}

Now, we return to the our problem where we have two dual Hamiltonian models where one is written in a computational basis and is a TC in uniform field and another one is written in the dual basis and is a transverse Ising model. Since the magnetization in Ising model is related to Pauli operator $\bar{X}_i$, one can ask which operator in the computational basis is mapped to an $\bar{X}_i$ operator in the dual basis. To this end, we consider TC on a 2D square lattice with an open boundary where plaquette operators in the boundary are three-local. As we explained in the previous section each operator $Z_e$ in the computational basis is mapped to an Ising interaction $\bar{X}_i \bar{X}_j$ in the dual basis. However, in TC with open boundary the situation is different for an edge qubit on the boundary. In other words, as it has been shown in Fig.(\ref{string}), the operator $Z_e$ which is applied in a qubit in the boundary does not commute only with one vertex operator of $A_0$. Therefore, it will be equal to a Pauli operator $\bar{X}_0$ in the dual basis. Therefore, the dual Hamiltonian is again a transverse Ising model with a little difference where in the boundary we have a single terms of $\bar{X}_0$ instead of Ising interactions. We should emphasize that this change in the boundary is not important in the phase transition because finally we will go to the thermodynamic limit.

\begin{figure}[t]
\centering
\includegraphics[width=9cm,height=5cm,angle=0]{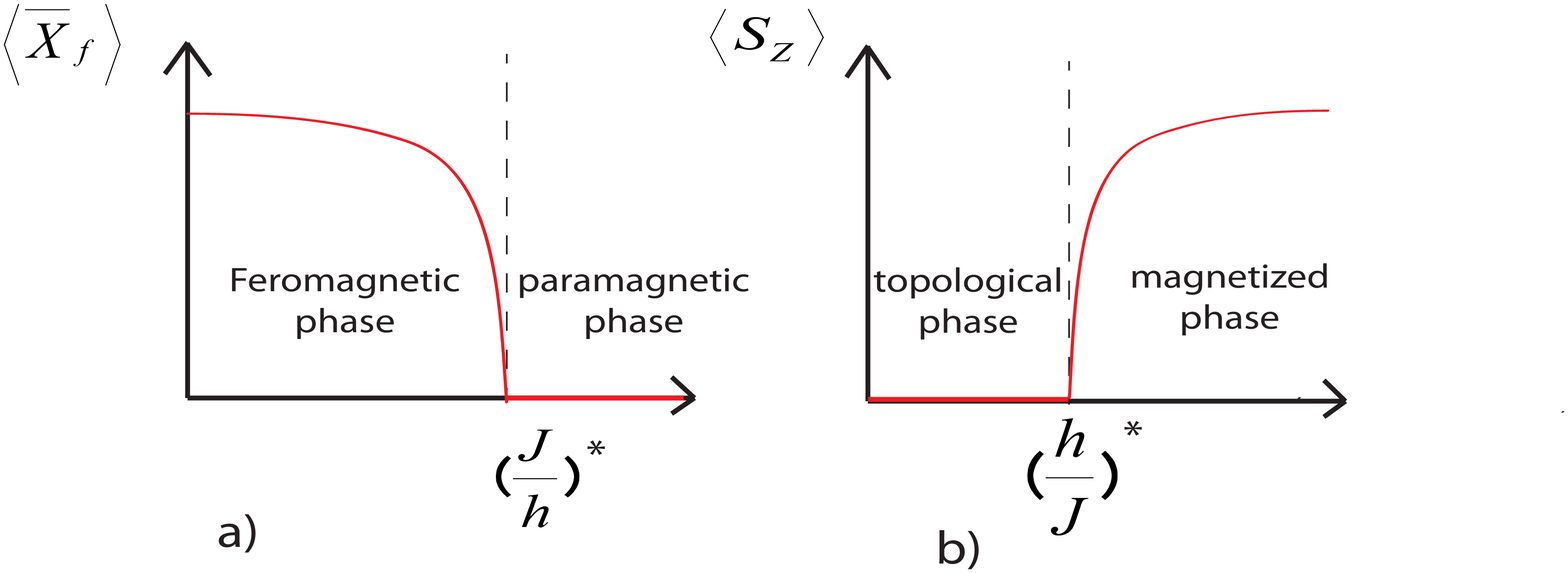}
\caption{(Color online) a) A schematic of phase diagram for the transverse Ising model Eq.(\ref{dual}) where $J$ refers to the magnetic term, $h$ refers to the Ising coupling and magnetization is the order parameter. b) A schematic of phase diagram for TC in magnetic field Eq.(\ref{TM}) where $h$ refers to the magnetic term, $J$ refers to TC coupling and the expectation value of the string operator of $S_Z$ plays the role of order parameter. Transition point is inverse of the transition point of the transverse Ising model in a sense that ferromagnetic phase is mapped to magnetized phase and paramagnetic phase is mapped to topological phase. } \label{order}
\end{figure}

In order to find dual of Ising order parameter, we consider a string in the lattice denoted by $S$ that one of its endpoints is in the boundary and another endpoint is in an arbitrary vertex $f$ as it has been shown in Fig.(\ref{string}). Corresponding to such a string, we define a string operator in the form of $S_z =\prod_{e\in S}Z_e$ where $e\in S$ refers to all edge qubits living in the string of $S$. Such string operator commutes with all vertex operators $A_i$ instead of one of them which is in the endpoint of the string i. e. $A_f$. Therefore and according to our previous arguments, the effect of this string operator in the dual basis is equal to a Pauli operator $\bar{X}_f$ which is applied to vertex qubit of $f$. On the other hand, since Ising order parameter is the expectation value of $\bar{X}_f$, it is concluded that Ising order parameter is equal to expectation value of string operator of $S_z$ in the TC model. It means that the expectation value of the string operator of $S_z$ is an order parameter which reveals topological phase transition in TC, see Fig.(\ref{order}). It is in particular interesting, because the string operator is non-local where if we go to the thermodynamic limit, it will be necessary to choose the vertex of $f$ far enough from the boundary where the length of the string should be scaled by the size of the lattice. Therefore, the string order parameter has a completely non-local nature which reveals topological nature of the quantum phase transition. On the other hand, the string order parameter has also an explicit topological nature because our mapping is independent of the geometry of the string where strings with the same endpoints lead to the same order parameter. Finally note that since the mapping is a strong-weak coupling duality, ferromagnetic and paramagnetic phases in the Ising model have been mapped to magnetized phase and topological phase in TC, respectively.

Finally, note that our the above argument will be held for TC on other graphs if we consider a suitable boundary condition. Therefore, since TC in uniform field on any graph is mapped to the transverse Ising model on the same graph, Ising order parameter is also mapped to string order parameter for TC on arbitrary graph. Consequently, the above string order parameter is a universal order parameter for TC model in uniform field.

\section{Discussion}
Topological phase transitions are different with ordinary quantum phase transitions because of their non-local nature. Therefore, it is expected that if there is an order parameter for a topological phase transition, it must be a non-local parameter. In this paper, we considered topological phase transition from TC states to a magnetized state. We introduced a string order parameter which could characterize topological nature of the above quantum phase transition. Specifically, we used a dual mapping where the above topological phase transition was mapped to the quantum phase transition in the transverse Ising model. Therefore, a symmetry-breaking phase transition with a local nature was used to characterize a topological phase transition with a non-local nature. \\
We would like also to emphasize in a speculative point as follows. TC model is the simplest model which describes a string-net condensed phase \cite{stringnet} where there is in fact a loop condensation in the ground state. On the other hand, quantum Ising model is also the simplest model which describes a particle condensed phase. Therefore, our mapping is interesting because it is in fact a mapping between two different physical mechanisms for symmetry-breaking and topological phase transition. The symmetry-breaking phase is described by a local (particle) order parameter and the topological phase is described by a non-local (string) order parameter. Therefore, it will be interesting to look for such mappings between string-net condensed phases and particle condensed phases in a more general way. It will be in fact a mapping between two different mechanisms of phase transitions.

\end{document}